\documentclass[11pt,aps,amssymb,superscriptaddress]{revtex4}

\usepackage{graphicx}
\usepackage[figuresright]{rotating}

\newcommand{\AmS}{{\protect\the\textfont2
  A\kern-.1667em\lower.5ex\hbox{M}\kern-.125emS}}

\begin{document}
\title{ Measuring the linear polarization of $\gamma$s in 20 - 170 GeV range
       }


\affiliation{Institute of Physics, Yerevan, Armenia}
\affiliation{Uppsala University, Uppsala, Sweden}
\affiliation{INFN and University of Firenze, Firenze, Italy}
\affiliation{INFN and University of Torino, Torino, Italy}
\affiliation{CERN, Geneva, Switzerland}
\affiliation{Northwestern University, Chicago,  USA}
\affiliation{INFN, Perugia, Italy}
\affiliation{Schonland Research Centre, Johannesburg, South Africa}
\affiliation{ESRF, Grenoble, France}
\affiliation{NIKHEF, Amsterdam, Netherlands}
\affiliation{Kurchatov Institute, Moscow, Russia}
\affiliation{University of Santiago de Compostela, Santiago de Compostela, Spain}
\affiliation{Institute of Nuclear Physics, Novosibirsk, Russia}
\affiliation{Uppsala University, Uppsala, Sweden}

\author{G.~Unel}
\email{gokhan.unel@fnal.gov}
\affiliation{Northwestern University, Chicago,  USA}

\author{A.~Apyan}
\affiliation{Institute of Physics, Yerevan, Armenia}

\author{R.O.~Avakian}
\affiliation{Institute of Physics, Yerevan, Armenia}

\author{B.~Badelek}
\affiliation{Uppsala University, Uppsala, Sweden}

\author{S.~Ballestrero}
\affiliation{INFN and University of Firenze, Firenze, Italy}

\author{C.~Biino}
\affiliation{INFN and University of Torino, Torino, Italy}
\affiliation{CERN, Geneva, Switzerland}

\author{I.~Birol}
\affiliation{Northwestern University, Chicago,  USA}

\author{P.~Cenci}
\affiliation{INFN, Perugia, Italy}

\author{S.H.~Connell}
\affiliation{Schonland Research Centre, Johannesburg, South Africa}

\author{S.~Eichblatt}
\affiliation{Northwestern University, Chicago,  USA}

\author{T.~Fonseca}
\affiliation{Northwestern University, Chicago,  USA}

\author{A.~Freund}
\affiliation{ESRF, Grenoble, France}

\author{B.~Gorini}
\affiliation{CERN, Geneva, Switzerland}

\author{R.~Groess}
\affiliation{Schonland Research Centre, Johannesburg, South Africa}

\author{K.~Ispirian}
\affiliation{Institute of Physics, Yerevan, Armenia}

\author{T.~Ketel}
\affiliation{NIKHEF, Amsterdam, Netherlands}

\author{Yu.V.~Kononets}
\affiliation{Kurchatov Institute, Moscow, Russia}

\author{A.~Lopez}
\affiliation{University of Santiago de Compostela, Santiago de Compostela, Spain}

\author{A.~Mangiarotti}
\affiliation{INFN and University of Firenze, Firenze, Italy}

\author{U.~Uggerhoj}
\affiliation{Institute for Storage Ring Facilities, University of Aarhus, Denmark}

\author{A.~Perego}
\affiliation{INFN and University of Firenze, Firenze, Italy}

\author{B.~van~Rens}
\affiliation{NIKHEF, Amsterdam, Netherlands}

\author{J.P.F.~Sellschop}
\thanks{Deceased}
\affiliation{Schonland Research Centre, Johannesburg, South Africa}

\author{M.~Shieh}
\affiliation{Northwestern University, Chicago,  USA}

\author{P.~Sona}
\affiliation{INFN and University of Firenze, Firenze, Italy}

\author{V.~Strakhovenko}
\affiliation{Institute of Nuclear Physics, Novosibirsk, Russia}

\author{E.~Uggerhoj}
\thanks{Co-Spokespersons}
\affiliation{Institute for Storage Ring Facilities, University of Aarhus, Denmark}

\author{M.~Velasco}
\thanks{Co-Spokespersons}
\affiliation{Northwestern University, Chicago,  USA}

\author{Z.Z.~Vilakazi}
\affiliation{Schonland Research Centre, Johannesburg, South Africa}

\author{U.~Wessley}
\affiliation{Uppsala University, Uppsala, Sweden}

\collaboration{The NA59 Collaboration}



\begin{abstract}
The Na59 collaboration aims to measure the linear polarization of its
photon beam in the 20-170 GeV range, using an aligned thin crystal. The 
tracks of $e^-/e^+$ pairs
created in two different crystal targets, germanium and diamond, are
reconstructed to obtain the photon spectrum. Using the polarization
dependence of the pair production cross section in an aligned crystal,  
photon polarization is obtained to be  55\% at the vicinity of 70 GeV.
\end{abstract}

\maketitle
\newpage
\section{Introduction}
The pair conversion in a thin crystal was proposed in the 1960s as a 
polarization measurement method for $\gamma$s in the few-GeV range
\cite{barbiellini}.
The fact that both the pair production cross section and the sensitivity to
$\gamma$ polarization increase with increasing $\gamma$ energy makes 
this method superior to others, such as the pair
production and photonuclear methods,
for present and future $\gamma$ beamlines.
The Na59 collaboration utilized it to map the polarization of its $\gamma$ beam.

A convenient way of creating a $\gamma$ beam with a predictable linear 
polarization spectrum is using the Coherent Bremsstrahlung (CB) 
\cite{cbdef} radiation from unpolarized electrons.
If the electron beam interacts coherently with the atoms 
in different planes in the crystal,  thus satisfying the Laue condition,
bremsstrahlung photons emerge with peaked energy values corresponding to
selected vectors of the reciprocal lattice. 
The energy and intensity  of these peaks are tunable by carefully 
aligning the lattice  planes with respect to the beam. 
In CB, the maximum of the polarization degree coincides with the maximum of
the intensity peak and polarizations up to 70\% have already been 
observed \cite{6gev} for 6 GeV electrons, and  up to 60\% for
higher energies \cite{omega}.

\begin{figure}[htb]
\begin{minipage}[t]{60mm} 
\vbox{
\includegraphics[scale=0.40]{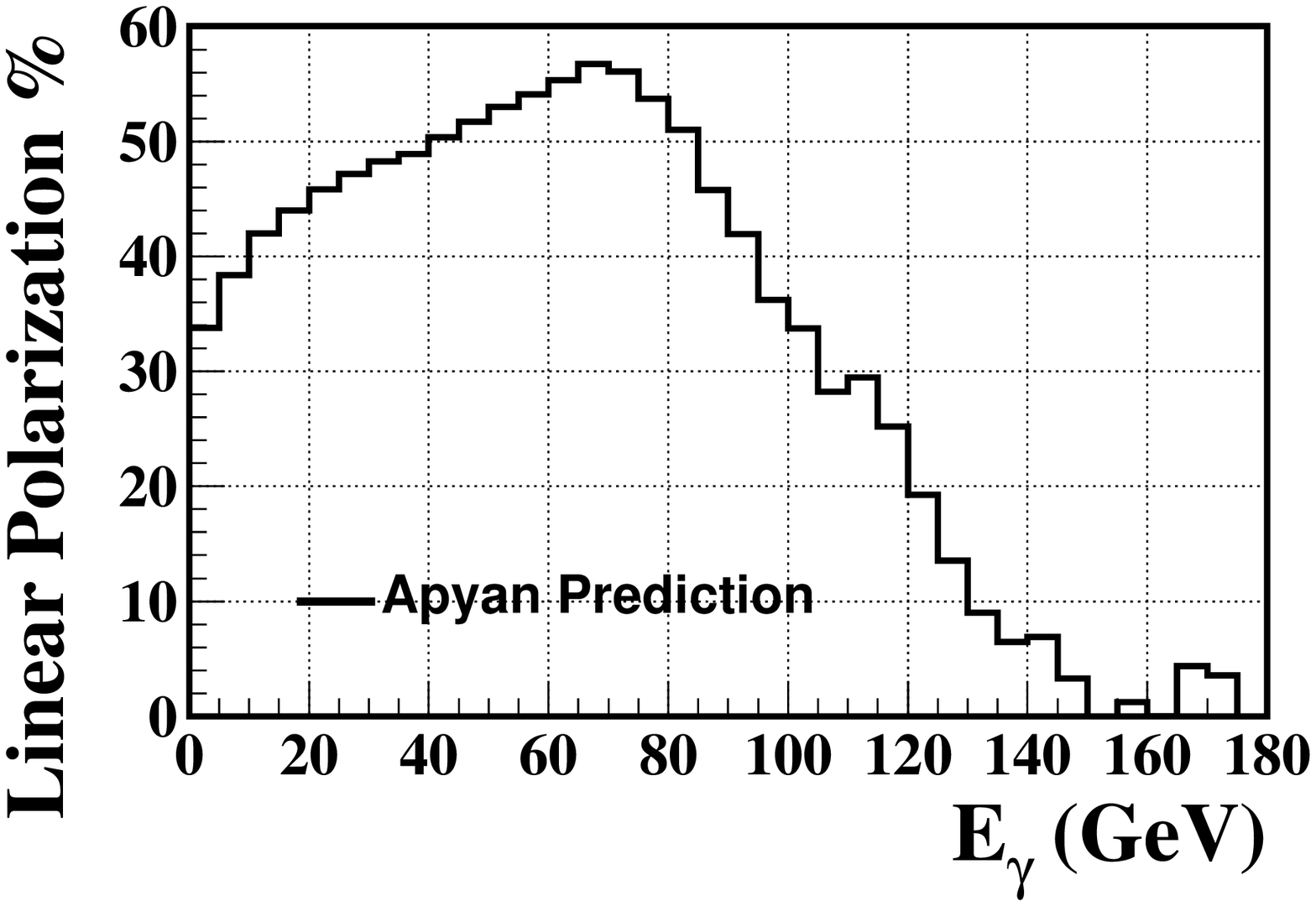}
\vskip-1cm}
\caption{Expected $\gamma$ polarization}
\label{fig:pol-predict}
\end{minipage}
\hspace{\fill}
\begin{minipage}[t]{80mm}
\vbox{
\includegraphics[scale=0.40]{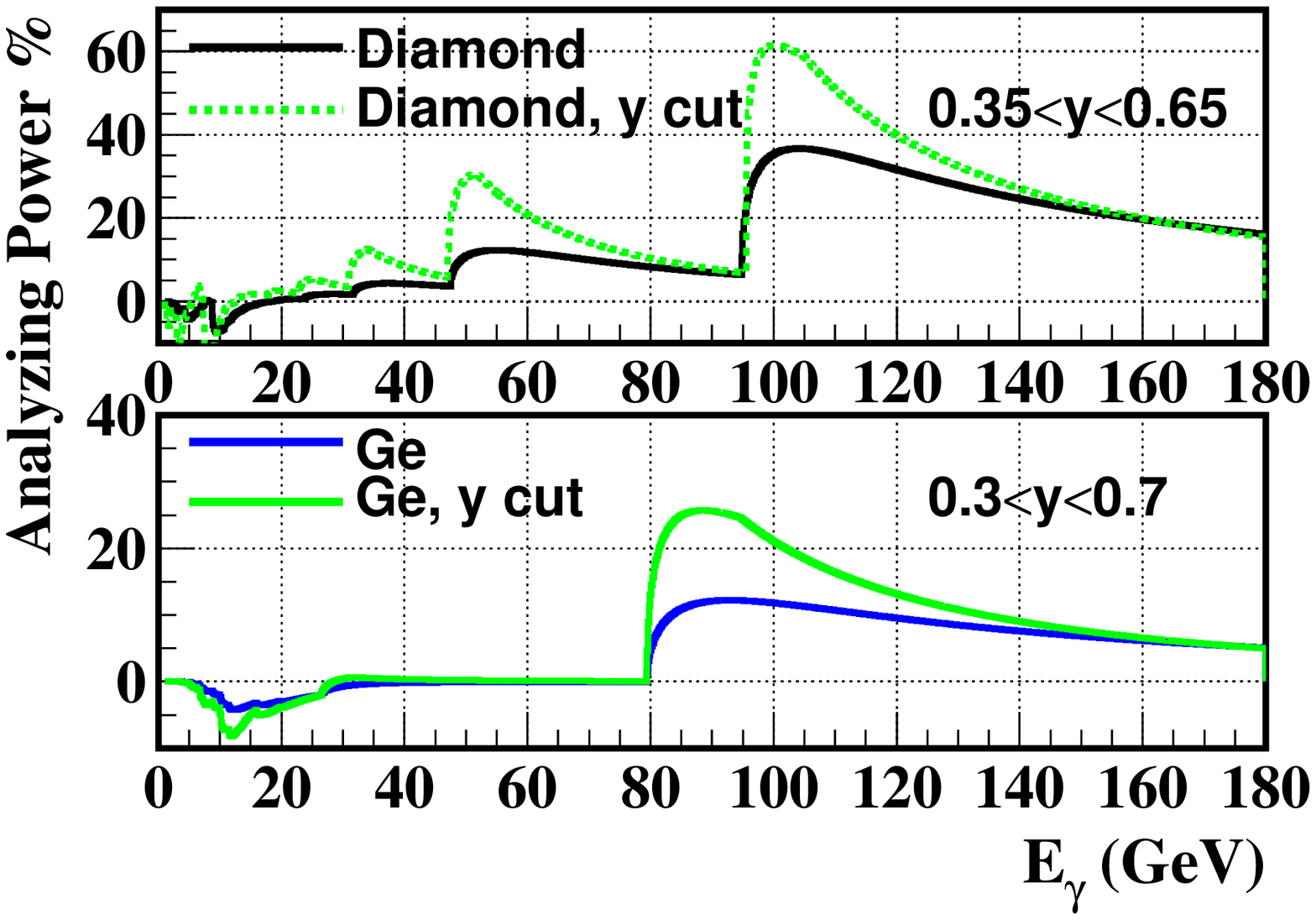}
\vskip-1cm}
\caption{$A$ for different targets 
($y\equiv E_e/E_\gamma$) }
\label{fig:ap}
\end{minipage}
\vskip-0.5cm
\end{figure}

The Na59 collaboration used 1.5cm thick Si single crystal with the electron 
beam making an angle of 5$m$rad from the $<$001$>$ crystallographic axis 
and about 70$\mu$rad from the (110) plane.
This choice yields a $\gamma$ beam with a maximum polarization of about 55\%
in the vicinity of 70 GeV, as can be seen from Figure \ref{fig:pol-predict}. This Monte Carlo calculation took into account the 
divergence (48$\mu$rad horizontally and 33$\mu$rad vertically) of the 
electron beam as well as its energy uncertainty of 1\%. The notation used 
is Stoke's polarization decomposition with Landau convention:
\begin{equation}
P_{\hbox {linear}}=\sqrt{\eta _{1}^{2}+\eta _{3}^{2}}\qquad P_{\hbox {circular}}=\sqrt{\eta _{2}^{2}}\qquad P_{\hbox {total}}=\sqrt{P_{\hbox {linear}}^{2}+P_{\hbox {circular}}^{2}} \quad .
\label{eq:pol-def}
\end{equation}
With the Na59 angular settings,  the photon polarization was solely created in 
the $\eta_{3}$ direction. We therefore made two distinct measurements, one of
$\eta_{3}$ to find the expected polarization, and another of $\eta_{1}$ to show
that it was consistent with zero.
The method for these two measurements is based on the birefringence properties
of the crystals. Since the imaginary part of the refraction index is 
proportional to the pair production probability, we defined
$\sigma_\parallel$ ($\sigma_\perp$) as the pair production cross
section when the selected crystallographic plane on the analyzer was 
parallel (perpendicular) to the photon polarization. The experimentally
relevant quantity is the asymmetry between these two cross sections and
it gives the $\gamma$ polarization, $\cal{P}$, through
\begin{equation}
a\equiv\frac{\sigma_\parallel-\sigma_\perp}
{\sigma_\parallel+\sigma_\perp}=A\times\cal{P} \quad ,
\label{eq:asym-def}
\end{equation}
where $A$ is the so called ``analyzing power'' of the crystal, and it 
represents the asymmetry for a 100\% polarized beam. A way of increasing
the analyzing power which can be computed accurately by MC techniques, is
to select the ``quasisymmetrical'' 
pairs \cite{ycut} in which the  $e^-$ and $e^+$
share the $\gamma$ energy almost equally. Figure \ref{fig:ap} shows the 
analyzing power with and without this selection (y cut) 
for two different analyzer crystal choices. 
This cut also reduces the relative statistical error 
on the asymmetry measured through the number of pair events in parallel
($N_\parallel$) and perpendicular ($N_\perp$) configurations : 
\begin{equation}
\frac{\delta a}{a} = \sqrt{ \frac{1-a^2}{a^2(N_\parallel +N_\perp)} } \quad .
\end{equation}

\section{Experimental Setup and Analysis}
\begin{figure}[htb]
\includegraphics[scale=0.56]{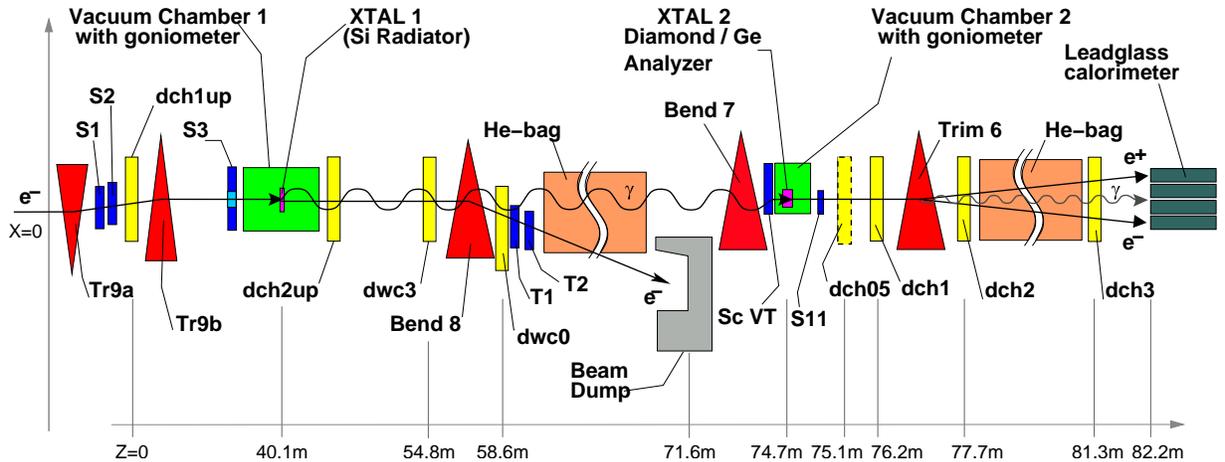}
\vskip-0.6cm
\caption{Experimental setup}
\label{fig:exp-setup}
\vskip-0.5cm
\end{figure}
The schematic view of the Na59 experimental setup is given in Figure 
\ref{fig:exp-setup}.
A tertiary 180 GeV electron beam from CERN SPS was sent onto the radiator 
crystal aligned with a goniometer of 2 $\mu$rad precision.  
Three upstream tracking chambers  defined
the entrance and exit angles of the electron beam.
The scattered electron beam is deflected toward the beam dump with a dipole magnet,
and passed through a tracking chamber to measure its  remaining energy.
The $\gamma$ beam is assumed to follow the direction of the 
incoming electrons impinged on the crystal target called the analyzer. 

The momenta of the pairs produced in the analyzer crystal are measured 
with a magnetic spectrometer consisting of a dipole magnet, 
two drift chambers (dch) downstream  and one  drift chamber 
upstream of it for the Ge target.
For the case of the multi-tile synthetic diamond target
\cite{luderitz}, a second dch (dch05) was
added right after it to improve the tracking.
The dch tracked charged particles with a resolution of 100$\mu m$. 
The total radiated energy was recorded with a 12 segment leadglass 
calorimeter with a resolution of $\frac{\sigma}{E}=11.5\% / \sqrt{E}$.

In the offline analysis, after applying beam quality cuts, the $e^-$ beam 
trajectory was found and the impact point on both radiator and analyzer 
crystals were determined for fiducial volume requirements. 
To reconstruct the photon energy in the pair spectrometer, an optimizing 
algorithm compensating for chamber inefficiencies and limited geometrical
acceptances was employed  \cite{swpap}.
The vertex reconstruction on the diamond analyzer allowed  veto of the
inter-tile events as well as the ones coming from a misaligned tile 
\cite{luderitz}. 

\section{Results and Conclusions}
To measure a polarization component, the asymmetry in Equation 
\ref{eq:asym-def} was experimentally constructed by taking data at 
two perpendicular analyzer crystal angular orientations.
After the mapping of both crystals was done, the data recording time for
each pair of angles  was two hours at the Na59 $e^-$ rate of 20KHz. 
To minimize the systematics, two measurements were performed with the
analyzer 180 degrees apart.  The measurement shown in 
Figure \ref{fig:false-asy} ensures that there is no ``false'' asymmetry
introduced due to analyzer crystal angular setting.
The zero asymmetry in Figure \ref{fig:eta1-rslt-ge} shows that
all linear polarization was in $\eta_3$ direction as expected. 
Figures \ref{fig:eta3-rslt-ge} and \ref{fig:eta3-rslt-di} show the measured 
asymmetries with and without the y cut for different analyzer crystals.
The asymmetries are in good agreement with theoretical predictions
in both cases.
In all Figures, the shaded region is the statistical error band
for the increase in asymmetry ($\Delta$asy) due to quasisymmetrical 
pair selection and it confirms the non statistical nature of the effect.
Comparing Figures \ref{fig:eta3-rslt-ge} and \ref{fig:eta3-rslt-di}, we 
conclude that multi-tile synthetic diamond is a better
choice than Ge as an analyzer, since for the 
same $\gamma$ polarization it yields a bigger asymmetry 
thus an easier measurement.

These results show that Na59 setup measures the polarization of 
high energy photons with good accuracy.
This measurement capability was used in other studies  \cite{newpaps} 
in Na59 research program, and will be reported elsewhere. 
We believe the presently investigated  crystal polarimetry technique 
is also applicable in future high energy photon beamlines as 
a fast monitoring tool.

\begin{center}
\vskip-0.25cm
\begin{figure}[htb]
\begin{minipage}[t]{70mm}
\includegraphics[scale=0.38]{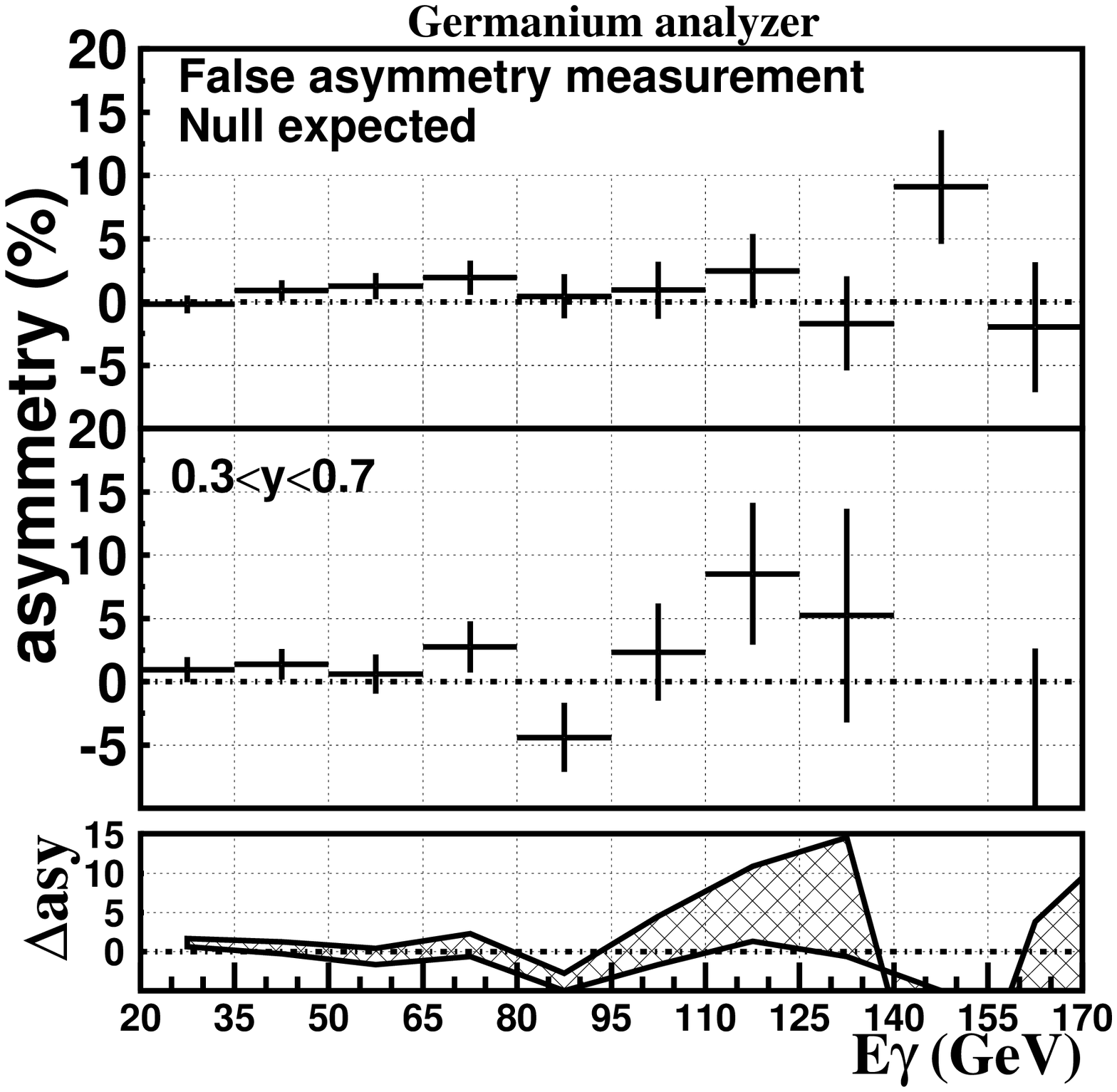}
\vskip-0.8cm
\caption{False asymmetry in Ge}
\label{fig:false-asy}
\end{minipage}
\hspace{\fill}
\begin{minipage}[t]{75mm}
\includegraphics[scale=0.38]{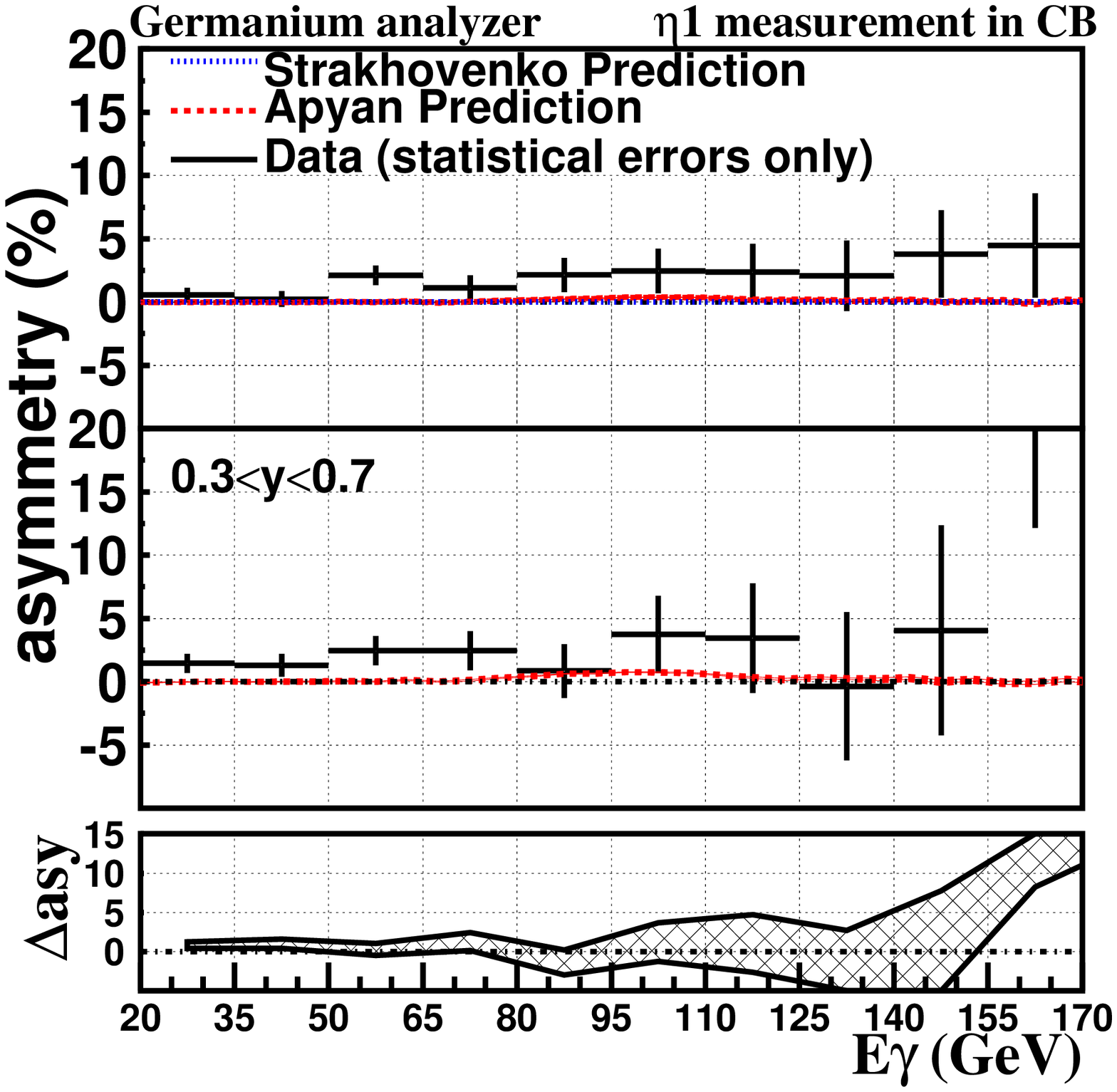}
\vskip-0.8cm
\caption{ $\eta_1$ measurement with Ge target}
\label{fig:eta1-rslt-ge}
\end{minipage}
\end{figure}
\vskip-1.5cm
\end{center}

\begin{center}
\begin{figure}[htb]
\begin{minipage}[t]{75mm}
\includegraphics[scale=0.38]{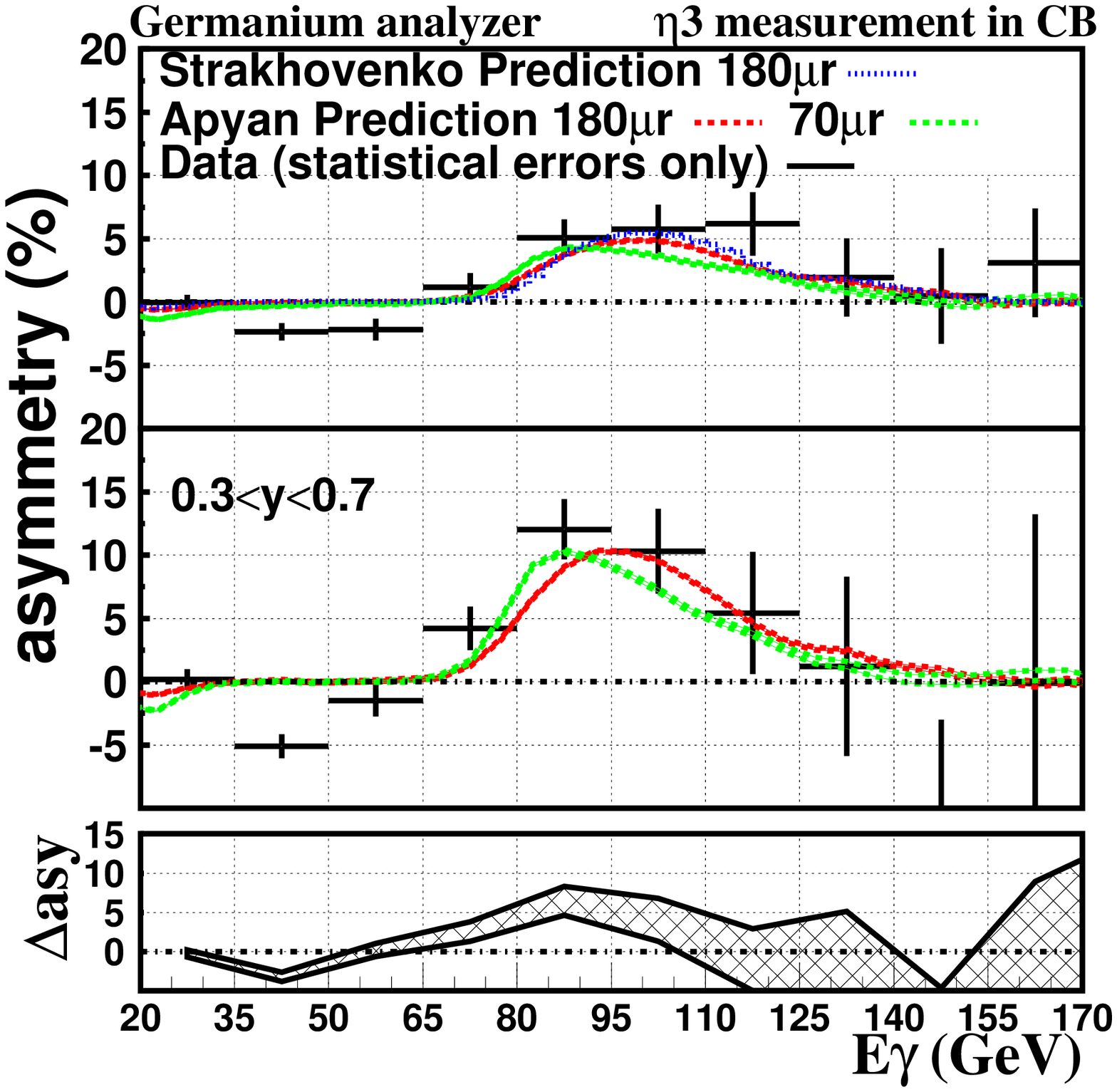}
\vskip-0.8cm
\caption{$\eta_3$ measurement with Ge target}
\label{fig:eta3-rslt-ge}
\end{minipage}
\hspace{\fill}
\begin{minipage}[t]{79mm}
\includegraphics[scale=0.38]{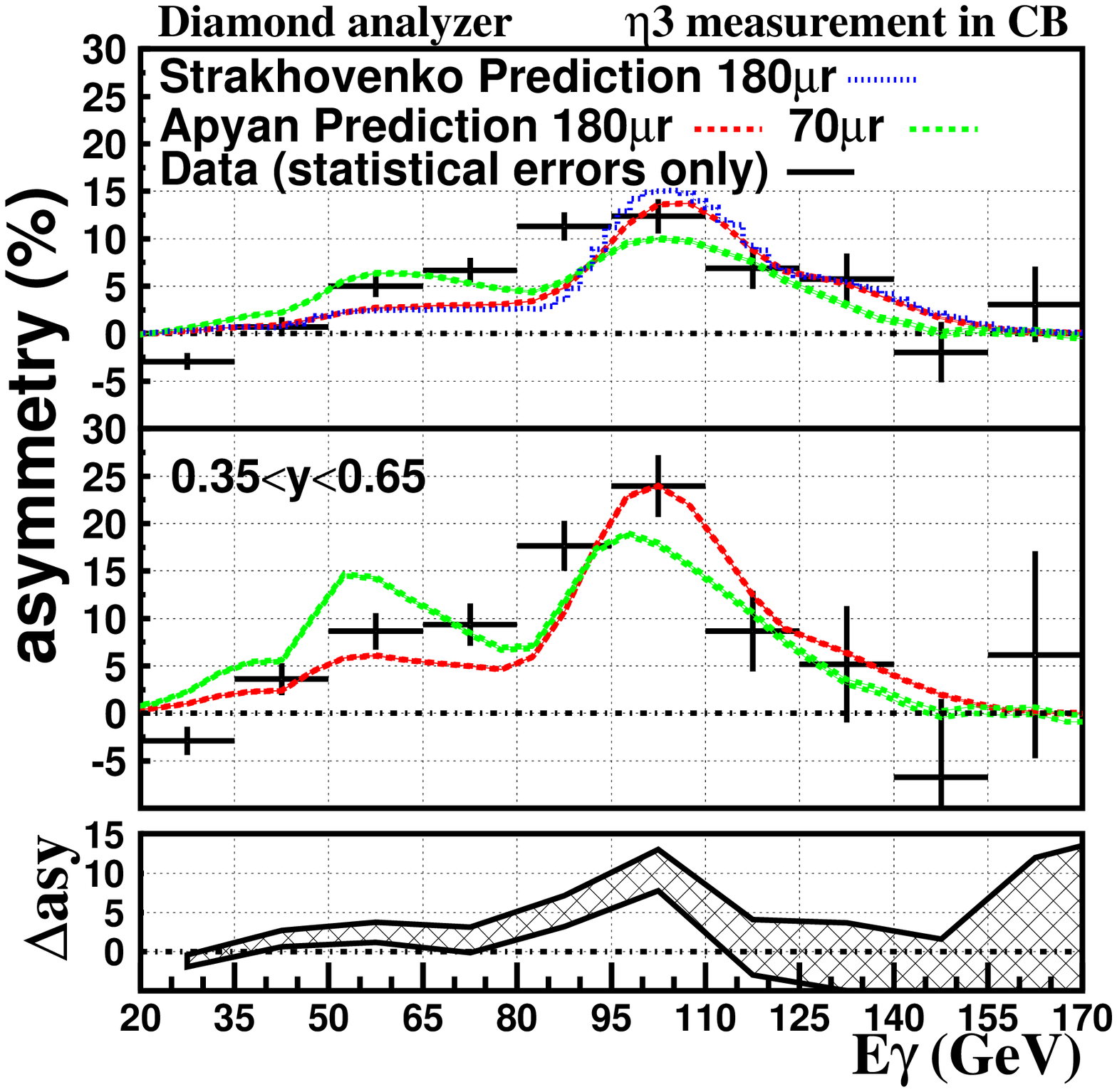}
\vskip-0.8cm
\caption{ $\eta_3$ measurement with diamond }
\label{fig:eta3-rslt-di}
\end{minipage}
\end{figure}
\vskip-1.3cm
\end{center}

\end{document}